\documentclass[english,aps,prl,reprint,superscriptaddress]{revtex4-1}
\usepackage[T1]{fontenc}
\usepackage[latin9]{inputenc}
\usepackage{color}
\usepackage{babel}
\usepackage{amsmath}
\usepackage{amssymb}
\usepackage{graphicx}
\usepackage[unicode=true,
 bookmarks=true,bookmarksnumbered=false,bookmarksopen=false,
 breaklinks=false,pdfborder={0 0 0},pdfborderstyle={},backref=false,colorlinks=true]
 {hyperref}
\hypersetup{
 allcolors=blue}

\makeatletter
\usepackage{braket}
\usepackage{dsfont}

\makeatother

\begin{document}
\title{Coherence dynamics induced by attenuation and amplification Gaussian
channels}
\author{Jonas F. G. Santos}
\email{jonasfgs18@gmail.com}

\affiliation{Centro de Ciências Naturais e Humanas, Universidade Federal do ABC,
Avenida dos Estados 5001, 09210-580 Santo André, São Paulo, Brazil}
\author{C. H. S. Vieira}
\email{vieira.carlos@ufabc.edu.br}

\affiliation{Centro de Ciências Naturais e Humanas, Universidade Federal do ABC,
Avenida dos Estados 5001, 09210-580 Santo André, São Paulo, Brazil}
\begin{abstract}
Quantum Gaussian channels play a key role in quantum information theory.
In particular, the attenuation and amplification channels are useful
to describe noise and decoherence effects on continuous variables
systems. They are directly associated to the beam splitter and two-mode
squeezing operations, which have operational relevance in quantum
protocols with bosonic models. An important property of these channels
is that they are Gaussian completely positive maps and the action
on a general input state depends on the parameters characterizing
the channels. In this work, we study the coherence dynamics introduced
by these channels on input Gaussian states. We derive explicit expressions
for the coherence depending on the parameters describing the channels.
By assuming a displaced thermal state with initial coherence as input
state, for the attenuation case it is observed a revival of the coherence
as a function of the transmissivity coefficient, whereas for the amplification
channel the coherence reaches asymptotic values depending on the gain
coefficient. Further, we obtain the entropy production for these class
of operations, showing that it can be reduced by controlling the parameters
involved. We write a simple expression for computing the entropy production
due to the coherence for both channels. This can be useful to simulate
many processes in quantum thermodynamics, as finite-time driving on
bosonic modes.
\end{abstract}
\maketitle

\section{Introduction}

In the last years, considerable effort has been dedicated to the comprehension
and application of bosonic continuous variables. The interest ranges
from quantum information theory \cite{Wang2007,Weedbrook2012,Adesso2014},
with direct implications in quantum communication and quantum computation
\cite{Marshall2019}, to quantum thermodynamics \cite{Macchiavello2020},
where bosonic systems may work as working substance in quantum heat
machines models \cite{Kosloff2017} and employed as ancillary systems
in structured thermal reservoir designs \cite{Chiara2018}. As in
general a system of interest is constantly interacting with its surrounding,
bosonic Gaussian channels are particularly powerful to describe noise
and decoherence in optical systems. There are two well-known bosonic
Gaussian channels, the attenuation and amplification channels \cite{Weedbrook2012,SerafiniBook},
as well as their limit cases, namely the quantum-limited attenuator
and the quantum-limited amplification. The concatenation of the last
two channels may be employed to represent any phase-insensitive channel
in quantum optic processes \cite{SerafiniBook}.

The actions of attenuation or amplification channels over a Gaussian
input state may be physically understood by the effect on the first
moments, resulting in a different output state. For the former, the
first moments of the output state are reduced depending on the value
of the transmissivity coefficient, whereas for the latter the output
state has its first moments enlarged given a value of gain coefficient.
Besides the emulation of noise and decoherence, the attenuation and
amplification channels are associated to the beam splitter and two-mode
squeezing operations \cite{Weedbrook2012,SerafiniBook}, respectively,
and thus directly related to the optical parametric down-conversion.
Moreover, the study of the capacity of Gaussian channels is an essential
task for any quantum communication protocol with continuous variables
\cite{Jarzyna2017,Qi2017}. For example, it has been shown that attenuation
and amplification channels transmit entanglement more efficiently
when the input state are non-Gaussian than Gaussian \cite{Filippov2014},
as well as they have been employed in continuous variable quantum-key
distribution \cite{Ruppert2019,Blandino2012} and quantum cloning
optimization \cite{Braunstein2001}. The connection of these type
of channels to non-Markovianity has been pointed in Ref. \cite{Liuzzo-Scorpo2017}.

The attenuation and amplification channels are well-known as Gaussian
completely positive (GCP) maps and they perform Gaussian operations
on input Gaussian states. On its turn, Gaussian operations are crucial
in many areas, such as quantum information science and quantum thermodynamics.
Depending on the process to be considered, Gaussian operations can
produce coherence on a given input state \cite{Aroch}. Recently,
coherence effects have been addressed in different branches of quantum
physics, for instance, in unitary process in quantum thermodynamics
where, in particular, assuming a driven Hamiltonian that does not
commute in different times will generate coherence in the system state
\cite{Camati2019}, as well as the investigation of coherence as a
resource in charging a quantum battery \cite{Garc=0000EDa-Pintos2020}.
In particular, the concept of quantum friction, directly associated
to the notion of finite-time regime in quantum protocols, has been
described in terms of coherence in a specific basis \cite{Rezek2010,Plastina2014}.
Moreover, some studies have associated an extra entropy production
cost to the presence of coherence in the state before the thermalization
\cite{Jader2019,Francica2019}. Besides coherence in quantum states,
the study of coherence in quantum channels has been investigated in
\cite{Datta2018,Xu2019}. Thus, controlling the coherence dynamics
of a specific protocol is certainly relevant, because the mitigation
of coherence could to optimize the energy cost of a process.

In this work, we are interested in studying the coherence dynamics
generated by the attenuation and amplification Gaussian channels over
input Gaussian states. Given the high precision control in quantum
optical platforms and its ability to generate these kind of Gaussian
channels, a detailed investigation of how they influence the initial
coherence of an input Gaussian state could be useful to design experiments
with bosonic Gaussian channels, as wells as their application in quantum
thermodynamics. Thus, we consider an input Gaussian state with initial
coherence whose dynamics is dictated by the attenuation (amplification)
channel. Employing a recent quantifier of coherence for Gaussian states,
we provide explicit expressions for the coherence dynamics for these
channels as functions of the parameters characterizing their actions.
Besides, motivated by recent investigations concerning the entropy
production in non-equilibrium quantum thermodynamics, we also study
the role performed by the attenuation (amplification) channel in the
entropy production. This could be useful for current and future experimental
realization in quantum thermodynamics and quantum information.

The manuscript is organized as follows. In section \ref{sec:Gaussian-states-and}
we review the main aspects and tools concerning the description of
$N$- mode Gaussian states and how to represent the attenuation and
amplification channel and parameterize them in terms of the beam splitter
and two-mode squeezing operations, respectively. We present our first
result in section \ref{sec:Coherence-in-attenuation}, where we derive
the coherence for a single-mode input Gaussian state passing through
the Gaussian channels referred above. Then, we generalize the results
by assuming a $N$- mode input Gaussian state in which each mode is
affected by an attenuation or amplification channel. Based on recent
advances in quantum thermodynamics, in section \ref{sec:Connection-to-quantum}
we study how the coherence dynamics introduced by the Gaussian channels
considered affects the entropy production during a thermalization
process. We derive analytical expression for that in terms of the
parameters characterizing the channels. Finally, in section \ref{sec:Conclusion}
we draw our conclusion and final remarks.

\section{Gaussian states and Gaussian channels\label{sec:Gaussian-states-and}}

In this section we introduce the basic tools concerning the notion
of Gaussian states and the class of phase-insensitive Gaussian channels,
and how to characterize them, depending on the action on input Gaussian
states.

Gaussian states have been largely applied in many branches of physics.
They are experimental accessed in platforms such as quantum optics
\cite{WallsBook} and trapped ions \cite{Ortiz-Guti=0000E9rrez2017}.
Their modern applications ranging from quantum information theory
\cite{Wang2007,Weedbrook2012,Adesso2014} to quantum thermodynamics
\cite{Macchiavello2020,Kosloff2017}. In order to characterize the
class of Gaussian states, we define the quadrature operators vector
$\vec{R}=\left(q_{1},p_{1},...q_{N},p_{N}\right)$, where $N$ is
the number of modes of a given system, and $\left(q,p\right)$ stand
for position and momentum, respectively. Gaussian states are completely
determined by the first moments and the covariance matrix, which are
defined as $\vec{d}=\langle\vec{R}\rangle_{\rho}$ and $\sigma_{ij}=\langle R_{i}R_{j}+R_{j}R_{i}\rangle_{\rho}-2\langle R_{i}\rangle_{\rho}\langle R_{j}\rangle_{\rho}$,
respectively, with $\rho=\rho(\vec{d},\sigma)$ the state of the system.
For a $N$-mode Gaussian state $\rho=\rho_{1}\otimes\rho_{2}\otimes...\otimes\rho_{N}$,
the associated first moments and covariance matrix are given by the
simple relations

\begin{align}
\vec{d} & =\vec{d}_{1}\oplus\vec{d}_{2}\oplus...\oplus\vec{d}_{N},\\
\sigma & =\sigma_{1}\oplus\sigma_{2}\oplus...\oplus\sigma_{N}.
\end{align}

The well-known example of single-mode Gaussian states is the thermal
state, with null first moments and covariance matrix given by $\sigma^{th}=(2\bar{n}+1)\mathbb{I}_{2}$,
with $\bar{n}=\text{Tr}\left[a^{\dagger}a\right]$ the average number
of photons in a mode, $a$ ($a^{\dagger}$) the annihilation (creator)
operator, and $\mathbb{I}_{2}$ an identity matrix of order 2. Thermal
states are written in the Fock basis as

\begin{equation}
\rho^{th}(\bar{n})=\sum_{n}\frac{\bar{n}^{n}}{\left(\bar{n}+1\right)^{n+1}}|n\rangle\langle n|
\end{equation}

Other examples, as the displaced thermal state, is obtained by acting
the Weyl operator on a general thermal state, while the coherent state
can be defined by the same action but with $\bar{n}=0$ \cite{WallsBook}.

Most part of the applications in quantum information theory and related
fields consider a dynamics where the system of interest (input state)
interacts with an ancillary one (ancilla) prepared in a Gaussian state,
also known as the environment. When a given protocol is implemented
through a quadratic Hamiltonian, then after tracing out the ancilla,
the Gaussianity of the initial input state is preserved \cite{SerafiniBook}.
The set of operations with this property are known as Gaussian completely
positive maps (GCP maps) or channels, being useful to describe noise
and decoherence effects on input Gaussian states. The complete characterization
of GCP maps on input Gaussian states can be represented by \cite{SerafiniBook}

\begin{align}
\vec{d} & \rightarrow\mathcal{F}\vec{d},\label{action1}\\
\sigma & \rightarrow\mathcal{F}\sigma\mathcal{F}^{T}+\mathcal{G},\label{action2}
\end{align}
 with $\mathcal{F}$ and $\mathcal{G}$ two $2N\times2N$ real matrices
satisfying the relation $\mathcal{G}+i\Omega\geq i\mathcal{F}\Omega\mathcal{F}^{T}$where
$\Omega=\left(\begin{array}{cc}
0 & 1\\
-1 & 0
\end{array}\right)$. Among different GCP maps which can be implemented, we are interested
in the well-known attenuation and amplification channels, also called
phase-insensitive Gaussian channels, which are responsible by many
optical processes of interest in quantum information theory \cite{Wang2007,Weedbrook2012,Adesso2014}.
The action of these channels on a single-mode input Gaussian state
can be written in terms of the quadrature operators vector. Following
Ref. \cite{Weedbrook2012} the attenuation channel is represented
by

\begin{equation}
\vec{R}\rightarrow\sqrt{\tau}\vec{R}+\sqrt{1-\tau}\vec{R}_{th},
\end{equation}
where $0<\tau<1$ is known as the transmissivity coefficient and $\vec{R}_{th}$
is the quadrature operator of the thermal state of the environment
with average number of photons $\bar{m}$. Similarly, the amplification
channel is given by

\begin{equation}
\vec{R}\rightarrow\sqrt{g}\vec{R}+\sqrt{g-1}\vec{R}_{th},
\end{equation}
 with $g>1$ the gain coefficient. We can parameterize these channels
by writing $\tau=\cos^{2}\theta$ and $g=\cosh^{2}r$ such that the
attenuation channel is identified to the action of a beam splitter,
whereas the amplification channel is identified to the action of a
two-mode squeezing transformation \cite{SerafiniBook}. In doing so,
we are able to represent the attenuation ($\mathcal{E}_{\theta}^{\bar{m}}$)
and the amplification ($\mathcal{A}_{r}^{\bar{m}}$) channels using
Eqs. (\ref{action1}) and (\ref{action2}) i.e.,

\begin{align}
\mathcal{E}_{\theta}^{\bar{m}} & :\label{attenuation}\\
\mathcal{F}_{\text{att}} & =\cos\theta\mathbb{I}_{2}\\
\mathcal{G}_{\text{att}} & =(2\bar{m}+1)\sin^{2}\theta\mathbb{I}_{2},
\end{align}
 with $\theta\in[0,2\pi[$, and

\begin{align}
\mathcal{A}_{r}^{\bar{m}} & :\label{amplification}\\
\mathcal{F}_{\text{amp}} & =\cosh r\mathbb{I}_{2}\\
\mathcal{G}_{\text{amp}} & =(2\bar{m}+1)\sinh^{2}r\mathbb{I}_{2},
\end{align}
with $r\in[0,\infty[$.

There are two limit cases which will be useful in the following, and
they are found by assuming $\bar{m}=0$. For the attenuation channel,
this results in an operation known as quantum limited attenuator.
In this limit, when the input state is a coherent state, the covariance
matrix remains unchanged. For the amplification channel, imposing
$\bar{m}=0$ implies in the channel known as quantum limited amplifier,
meaning that, a minimum noise is added in the covariance matrix of
the input state. It is possible to show that any single-mode phase-insensitive
Gaussian channel can be written using the quantum limited attenuator
and the quantum limited amplifier channels \cite{SerafiniBook}, i.e,

\begin{align}
\mathcal{E^{\text{ph}}} & =\mathcal{A}_{r}^{0}\circ\mathcal{E}_{\theta}^{0}:\label{phaseinsensitive}\\
\mathcal{F}_{\text{ph}} & =\cosh r\cos\theta\mathbb{I}_{2}\\
\mathcal{G}_{\text{ph}} & =\left(\cosh^{2}r\sin^{2}\theta+\sinh^{2}r\right)\mathbb{I}_{2}.
\end{align}

This channel is particularly important to study the information capacity
of bosonic channels \cite{SerafiniBook}.

\section{Coherence in attenuation and amplification channels\label{sec:Coherence-in-attenuation}}

The investigation of coherence in quantum systems is an active field
in quantum information theory \cite{Plenio2014,Winter2016,Dana2017},
quantum thermodynamics \cite{Macchiavello2020} and related areas
\cite{Ralph2003,Joo2011}. Given an initial state $\rho_{0}$ subject
to a given protocol represented by a map $\Lambda_{\tau,0}$, the
ability to quantify the coherence of the final state $\rho_{\tau}=\Lambda_{\tau,0}\left[\rho_{0}\right]$
is an important task. Since the coherence depends on the particular
choice of a basis, we use here the Fock basis $\left\{ |N\rangle\right\} _{0}^{\infty}$,
which has the advantage to be associated to the eigenvalues of energy
of a system. Following Ref. \cite{Plenio2014}, to introduce an entropic
quantifier of coherence we must define a set of incoherent states.
From Ref. \cite{Xu2016}, for general Gaussian states the set of incoherent
states are given by the thermal states, which are Gibbs states and
do not present coherence in the Fock basis. By defining a reference
state as being a thermal one $\zeta=\rho^{th}(\bar{k})$ we can write
an entropic quantifier of coherence of a Gaussian state $\rho$ as
$C(\rho)=\text{min}_{\zeta}\left[S\left(\rho||\zeta\right)\right]$
with $S\left(\rho_{1}||\rho_{2}\right)=\text{Tr\ensuremath{\left[\rho_{1}\text{ln}\rho_{1}\right]}}-\text{Tr}\left[\rho_{1}\text{ln}\rho_{2}\right]$
the relative entropy and the minimization is performed on all the
thermal states $\zeta$.

Following Ref. \cite{Xu2016}, an entropic measure for coherence of
a $N$-mode Gaussian state $\rho=\rho(\vec{d},\sigma)$ is given by

\begin{equation}
C(\rho)=S(\zeta)-S(\rho),\label{coherencemeasure}
\end{equation}
 where $S(\cdot)$ is the von Neumann entropy,

\begin{equation}
S(\rho)=-\sum_{j=1}^{N}\left[\frac{\nu_{j}-1}{2}\ln\left(\frac{\nu_{j}-1}{2}\right)-\frac{\nu_{j}+1}{2}\ln\left(\frac{\nu_{j}+1}{2}\right)\right],
\end{equation}
 with $\left\{ \nu_{j}\right\} _{j=1}^{N}$ the symplectic eigenvalues
of $\sigma$, and $\zeta$ is a $N$-mode reference thermal state
with average number of photons $\left\{ \bar{k}_{j}\right\} _{j=1}^{N}$written
in terms of the first moments and the covariance matrix of the input
state and given by

\begin{equation}
\bar{k}_{j}=\frac{1}{4}\left[\sigma_{11}^{j}+\sigma_{22}^{j}+\left(d_{1}^{j}\right)^{2}+\left(d_{2}^{j}\right)^{2}-2\right].
\end{equation}

The coherence quantifier in Eq. (\ref{coherencemeasure}) satisfies
the following properties: $C(\rho)\geq0$, $C(\rho)=0$ if and only
if $\rho$ is a tensor product of thermal states, $C(\rho)\geq C(\Phi_{IGC}\rho)$,
where $\Phi_{IGC}$ is an incoherent Gaussian channel \cite{Xu2016}.
In the following, we consider the problem of a single-mode input Gaussian
state. Then, we generalize our results for a $N$-mode input Gaussian
state.

\subsection*{One-mode input state}

Consider a single-mode input Gaussian state $\rho=\rho(\vec{d},\sigma)$
passing through a Gaussian channel as depicted in Fig. 1-a). The channel
can be an attenuation or an amplification one, and the ancilla is
assumed to be a Gaussian state with average number of photons $\bar{m}$
in both cases. We consider that the single-mode input state is Gaussian,
more specifically, it is a displaced thermal state with first moments
$\vec{d}=\left(q_{0},p_{0}\right)$ and covariance matrix $\sigma=\left(2\bar{n}+1\right)\mathbb{I}_{2}$.
Using Eq. (\ref{coherencemeasure}) for the attenuation channel introduced
in Eq. (\ref{attenuation}), the coherence quantifier can be written
in a compact form as

\begin{align}
C_{\text{att}}(\rho) & =\frac{A_{\text{att}}-1}{2}\ln\left[\frac{A_{\text{att}}-1}{2}\right]\nonumber \\
 & -\frac{A_{\text{att}}+1}{2}\ln\left[\frac{A_{\text{att}}+1}{2}\right]\nonumber \\
 & +B_{\text{att}}\ln\left[B_{\text{att}}\right]-\left(B_{\text{att}}-1\right)\ln\left[B_{\text{att}}-1\right],\label{Att}
\end{align}

where

\begin{align}
A_{\text{att}} & =\left(2\bar{n}+1\right)\cos^{2}\theta+(2\bar{m}+1)\sin^{2}\theta,\\
B_{\text{att}} & =\frac{1}{2}\left(\left(2\bar{n}+1\right)\cos^{2}\theta+\left(2\bar{m}+1\right)\sin^{2}\theta\right)\\
 & +\frac{1}{4}\left[\left(q_{0}^{2}+p_{0}^{2}\right)\cos^{2}\theta-2\right]+1.
\end{align}

For the amplification channel, a similar calculation provides the
following expression for the coherence

\begin{align}
C_{\text{amp}}(\rho) & =\frac{A_{\text{amp}}-1}{2}\ln\left[\frac{A_{\text{amp}}-1}{2}\right]\nonumber \\
 & -\frac{A_{\text{amp}}+1}{2}\ln\left[\frac{A_{\text{amp}}+1}{2}\right]\nonumber \\
 & +B_{\text{amp}}\ln\left[B_{\text{amp}}\right]-\left(B_{\text{amp}}-1\right)\ln\left[B_{\text{amp}}-1\right]\label{Amp}
\end{align}

where

\begin{align}
A_{\text{amp}} & =\left(2\bar{n}+1\right)\cosh^{2}r+(2\bar{m}+1)\sinh^{2}r,\\
B_{\text{amp}} & =\frac{1}{2}\left(\left(2\bar{n}+1\right)\cosh^{2}r+\left(2\bar{m}+1\right)\sinh^{2}r\right)\\
 & +\frac{1}{4}\left[\left(q_{0}^{2}+p_{0}^{2}\right)\cosh^{2}r-2\right]+1.
\end{align}

In order to illustrate the coherence dynamics introduced by the attenuation
and amplification channels on the single-mode input Gaussian state,
we consider the initial first moments to be $\left(q_{0},p_{0}\right)=\left(1,1\right)$
in arbitrary unity, an average number of photons $\bar{n}=4$ for
the input state. Figure \ref{cohereces1mode} illustrates the coherence
quantifier for the attenuation (figure a)) and the amplification (figure
b)) channels, as a function of the parameter $\theta$ and $r$, respectively.

\begin{figure}
\includegraphics[scale=0.1]{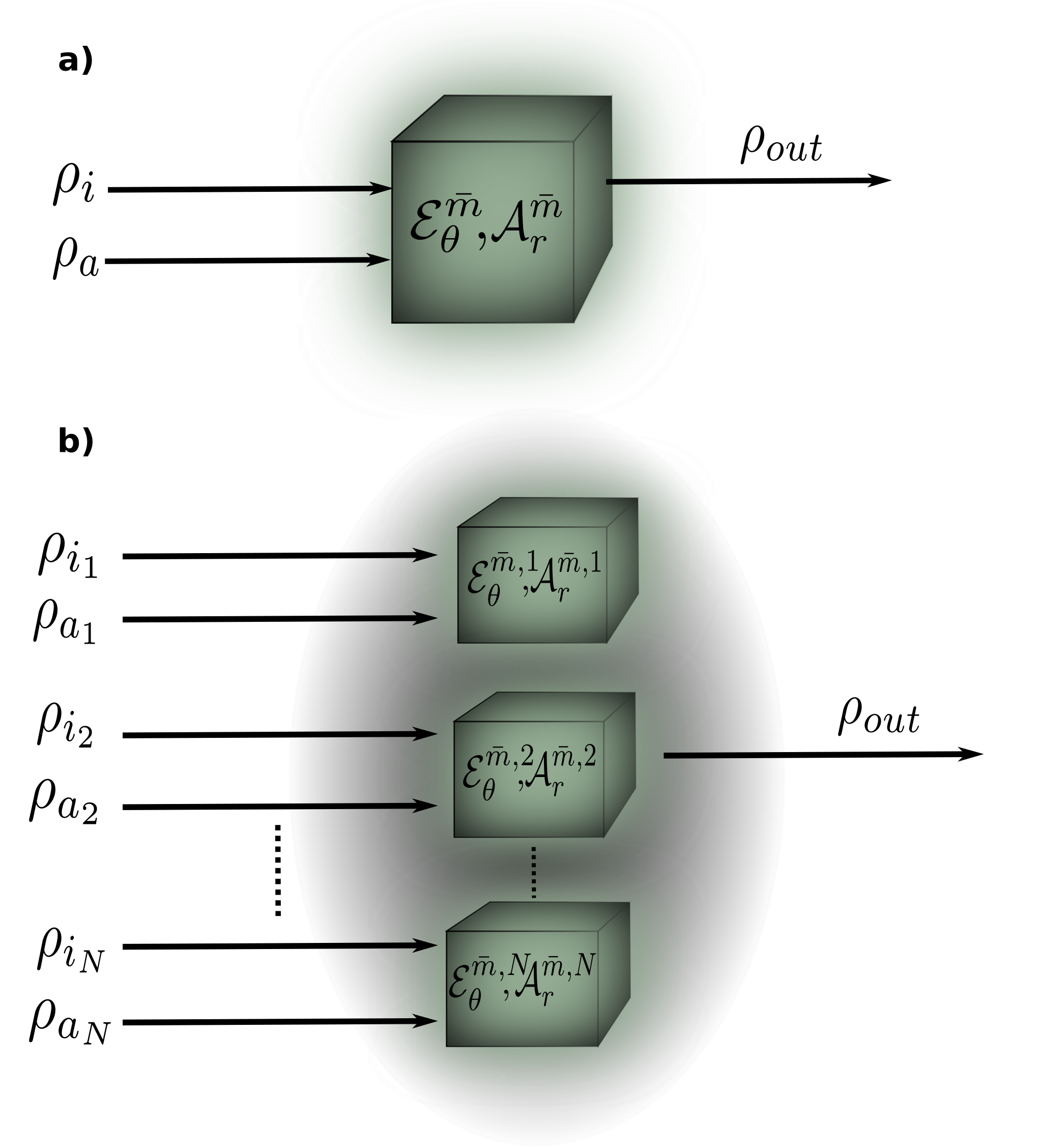}

\label{Fggg}\caption{Action of Gaussian channels. a) A single-mode input Gaussian state
passes through an attenuation (amplification) channel parameterized
by the parameters $\bar{m}$ and $\theta$ ($\bar{m}$ and $r$).
Since the input state is assumed to have an initial amount of coherence,
the coherence dynamics of the output state will depend on the Gaussian
channel considered. b) Generalization to a $N$-mode input Gaussian
state where each mode interacts with an attenuation (amplification)
channel separately.}
\end{figure}

\begin{figure}
\includegraphics[scale=0.7]{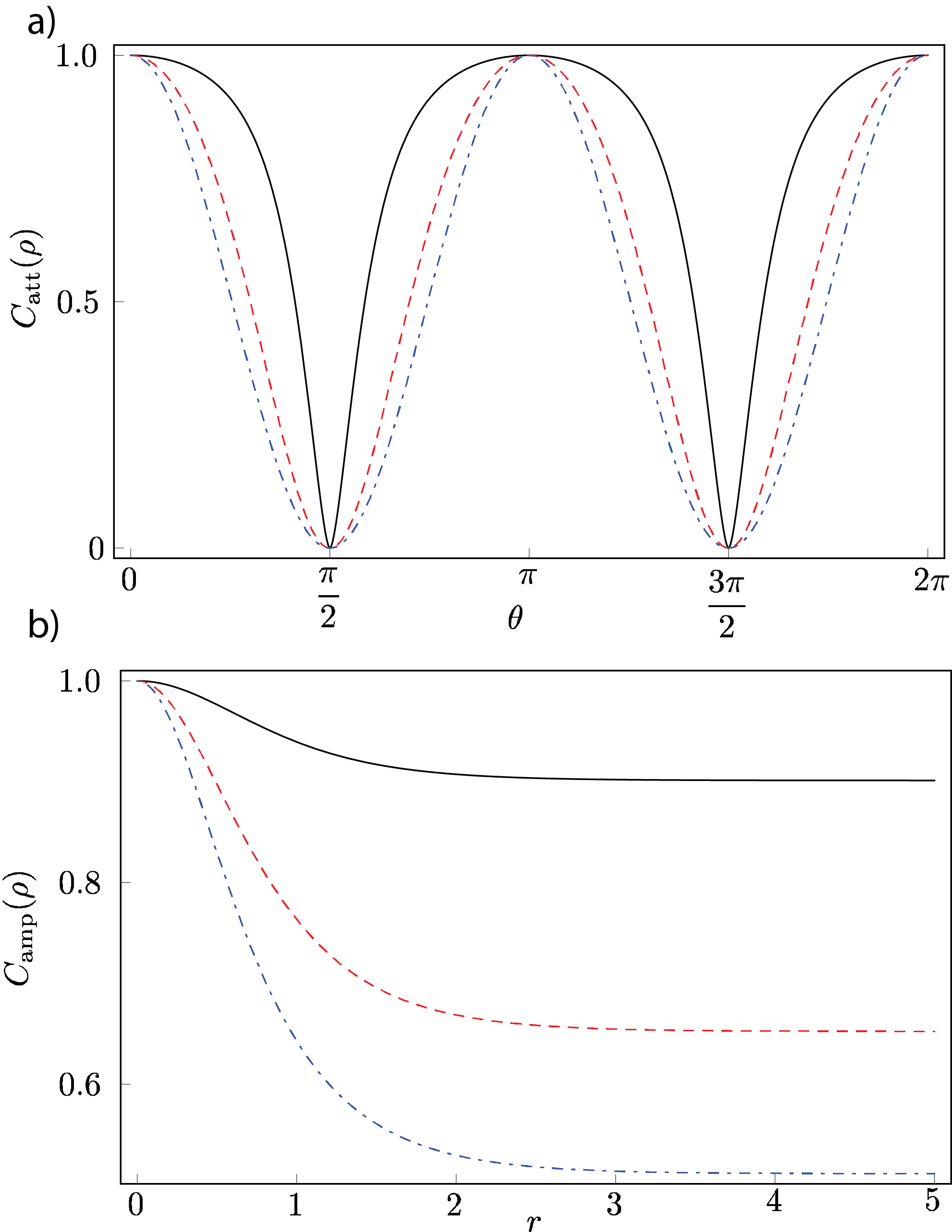}

\caption{Normalized coherence quantifier of the output single-mode Gaussian
state. a) Coherence of the output Gaussian state due to the attenuation
channel. b) Coherence of the output Gaussian state due to the amplification
channel. We consider an average number of photons $\bar{n}=4$ and
initial first moments $(q_{0},p_{0})=(1,1)$ for the input state.
For the thermal environment, we assume $\bar{m}=0$ (solid black line),
$\bar{m}=2$ (red dashed line), and $\bar{m}=4$ (blue dotted dashed
line). For $\bar{m}=0$ we have the quantum limited attenuator and
the quantum limited amplifier, respectively.}

\label{cohereces1mode}
\end{figure}

Figure \ref{cohereces1mode}-a)\textbf{ }shows that for the attenuation
channel, the coherence of the single-mode input Gaussian state oscillates
between its initial value and zero. Indeed, when $\theta=\pi/2$ and
$3\pi/2$ the first moments of the output state are zero, and the
input state, which was assumed to be a displaced thermal state, becomes
a purely thermal one. It is possible also to note that the quantum
limited attenuator, $\bar{m}=0$ (solid black line), represents the
smoothest value of coherence at each maximum point. This could be
interesting in protocols where the aim is to preserve the coherence
in a wide range of a given parameter.

On the other hand, for the amplification Gaussian channel, Fig. \ref{cohereces1mode}-b)\textbf{
}depicts the coherence of the single-mode output state as a function
of the two-mode squeezing operation parameter $r$. We observe that,
as expected, the quantum limited amplifier case, $\bar{m}=0$ (solid
black line) means the minimum influence on the covariance matrix of
the input state, implying in the maximum value of coherence in the
asymptotic limit. At first sight, we could expect that the coherence
introduced by the amplification channel would increase as the parameter
$r$ becomes larger, by virtue of, for Gaussian states, the coherence
increases as more distant the state is from the origin on the phase-space.
However, we must take in account the action of the amplification channel
on the covariance matrix of the input state, which is responsible
for the behavior shown in Fig. \ref{cohereces1mode}-b). From Eq.
(\ref{Amp}) we are able to obtain the asymptotic value of coherence
when $r\rightarrow0$ and $r\rightarrow\infty$. They are given respectively
by

\begin{align}
C_{\text{amp}}^{r\rightarrow0}(\rho) & =\bar{n}\ln\left(\bar{n}\right)-\left(\bar{n}+1\right)\ln\left(\bar{n}+1\right)\nonumber \\
 & +\left(\bar{n}+1+\frac{q_{0}^{2}+p_{0}^{2}}{4}\right)\ln\left(q_{0}^{2}+p_{0}^{2}+4\bar{n}+4\right)\nonumber \\
 & -\left(\bar{n}+\frac{q_{0}^{2}+p_{0}^{2}}{4}\right)\ln\left(q_{0}^{2}+p_{0}^{2}+4\bar{n}\right)-2\ln(2),\label{Limit01}
\end{align}

and

\begin{align}
C_{\text{amp}}^{r\rightarrow\infty}(\rho) & =\ln\left(q_{0}^{2}+p_{0}^{2}+4\left(\bar{n}+\bar{m}+1\right)\right)\nonumber \\
 & -\ln\left(\bar{n}+\bar{m}+1\right)-2\ln(2).\label{Limit02}
\end{align}

Equation (\ref{Limit01}) is nothing but the initial coherence of
the single-mode input Gaussian state. On the other hand, Eq. (\ref{Limit02})
represents the maximum coherence that one achieves in the output state,
despite the fact that its first moments increase as the parameter
$r$ becomes larger. Another important limit we can investigate is
when the average number of photons of the input state, $\bar{n}$,
goes to infinity, i.e., the input state possesses considerable thermal
fluctuation effects. In this case, Eqs. (\ref{Att}) and (\ref{Amp})
result in $C_{\text{att}}^{\bar{n}\rightarrow\infty}=C_{\text{amp}}^{\bar{n}\rightarrow\infty}=0$,
indicating that, if the input Gaussian state is highly thermal, then
no coherence dynamics is introduced by the attenuation and the amplification
channels. This is similar to the thermodynamic limit, in which $k_{B}T\gg\hbar\omega$.

To conclude our discussion about single-mode input states, we consider
the concatenation of channels defined in Eq. (\ref{phaseinsensitive}).
It is straightforward to note that the coherences displayed in Figs.
\ref{cohereces1mode}-a)\textbf{ }and \ref{cohereces1mode}-b)\textbf{
}are marginals of the concatenation of the quantum limited attenuator
and the quantum limited amplifier, i.e., the particular case of $\bar{m}=0$
(solid black lines).

\subsection*{N-mode input state}

In order to generalize our results, we consider now a $N$-mode input
Gaussian state formed by the tensor product of single-mode states,
given by $\rho=\rho(\vec{d},\sigma)$, with $\vec{d}=\left(q_{1,0},p_{1,0},...,q_{N,0},p_{N,0}\right)$
and $\sigma=\sigma_{1,0}\oplus\sigma_{2,0}\oplus...\oplus\sigma_{N,0}$
the initial first moments and covariance matrix, respectively, and
$\sigma_{j,0}=\left(2\bar{n}_{j}+1\right)\mathbb{I}_{2}$. The state
$\rho$ passes through a GCP map composed of a series of independent
attenuation (amplification) channels $\mathcal{E}_{\theta}^{\bar{m},j}$($\mathcal{A}_{r}^{\bar{m},j}$),
as depicted in Fig. 2-b).

The coherence quantifier in this situation is a straightforward generalization
of the one-mode case and, for the attenuation channel, it is given
by

\begin{align}
C_{\text{att}}(\rho) & =-S(\rho)+\sum_{j=1}^{N}\left[\left(\bar{k}_{j}+1\right)\ln\left(\bar{k}_{j}+1\right)-\left(\bar{k}_{j}\right)\ln\left(\bar{k}_{j}\right)\right],\label{NAtt}
\end{align}
where

\begin{equation}
S(\rho)=-\sum_{j=1}^{N}\left[\frac{\nu_{j}-1}{2}\ln\left(\frac{\nu_{j}-1}{2}\right)-\frac{\nu_{j}+1}{2}\ln\left(\frac{\nu_{j}+1}{2}\right)\right],
\end{equation}
 with $\nu_{j}$ the symplectic eigenvalues of the output covariance
matrix, given by

\begin{equation}
\nu_{j}=\left(2\bar{n}_{j}+1\right)\cos^{2}\theta+(2\bar{m}_{j}+1)\sin^{2}\theta,
\end{equation}
 and $\bar{k}_{j}$ is the $j$-th average number of photons associated
to the $j$-th thermal reference state, and written as

\begin{align*}
\bar{k}_{j} & =\frac{1}{2}\left(\left(2\bar{n}_{j}+1\right)\cos^{2}\theta+(2\bar{m}_{j}+1)\sin^{2}\theta\right)\\
 & +\frac{1}{4}\left(\cos^{2}\theta\left(q_{j,0}^{2}+p_{j,0}^{2}\right)-2\right),
\end{align*}
whereas for the amplification channel one has

\begin{align}
C_{\text{amp}}(\rho) & =-S(\rho)+\sum_{j=1}^{N}\left[\left(\bar{k}_{j}+1\right)\ln\left(\bar{k}_{j}+1\right)-\left(\bar{k}_{j}\right)\ln\left(\bar{k}_{j}\right)\right],\label{NAmp}
\end{align}
where

\begin{equation}
S(\rho)=-\sum_{j=1}^{N}\left[\frac{\nu_{j}-1}{2}\ln\left(\frac{\nu_{j}-1}{2}\right)-\frac{\nu_{j}+1}{2}\ln\left(\frac{\nu_{j}+1}{2}\right)\right],
\end{equation}
 with $\nu_{j}$ the symplectic eigenvalues of the output covariance
matrix, given by

\begin{equation}
\nu_{j}=\left(2\bar{n}_{j}+1\right)\cosh^{2}r+(2\bar{m}_{j}+1)\sinh^{2}r,
\end{equation}
 and $\bar{k}_{j}$ is the $j$-th average number of photons associated
to the $j$-th thermal reference state, and written as

\begin{align*}
\bar{k}_{j} & =\frac{1}{2}\left(\left(2\bar{n}_{j}+1\right)\cosh^{2}r+(2\bar{m}_{j}+1)\sinh^{2}r\right)\\
 & +\frac{1}{4}\left(\cosh^{2}r\left(q_{j,0}^{2}+p_{j,0}^{2}\right)-2\right).
\end{align*}

Equations (\ref{NAtt}) and (\ref{NAmp}) are general in the sense
that each attenuation (amplification) channel acting on each mode
of the input state are free to have different average number of photons
$\bar{m}_{j}$. Besides, although we have considered the same value
of $\theta$ and $r$ for different $j$-th attenuation (amplification)
channel, the expressions could be easily generalized for different
values, corresponding to $\theta_{j}$ and $r_{j}$.

\section{Connection to quantum thermodynamics\label{sec:Connection-to-quantum}}

\begin{figure}
\includegraphics[scale=0.06]{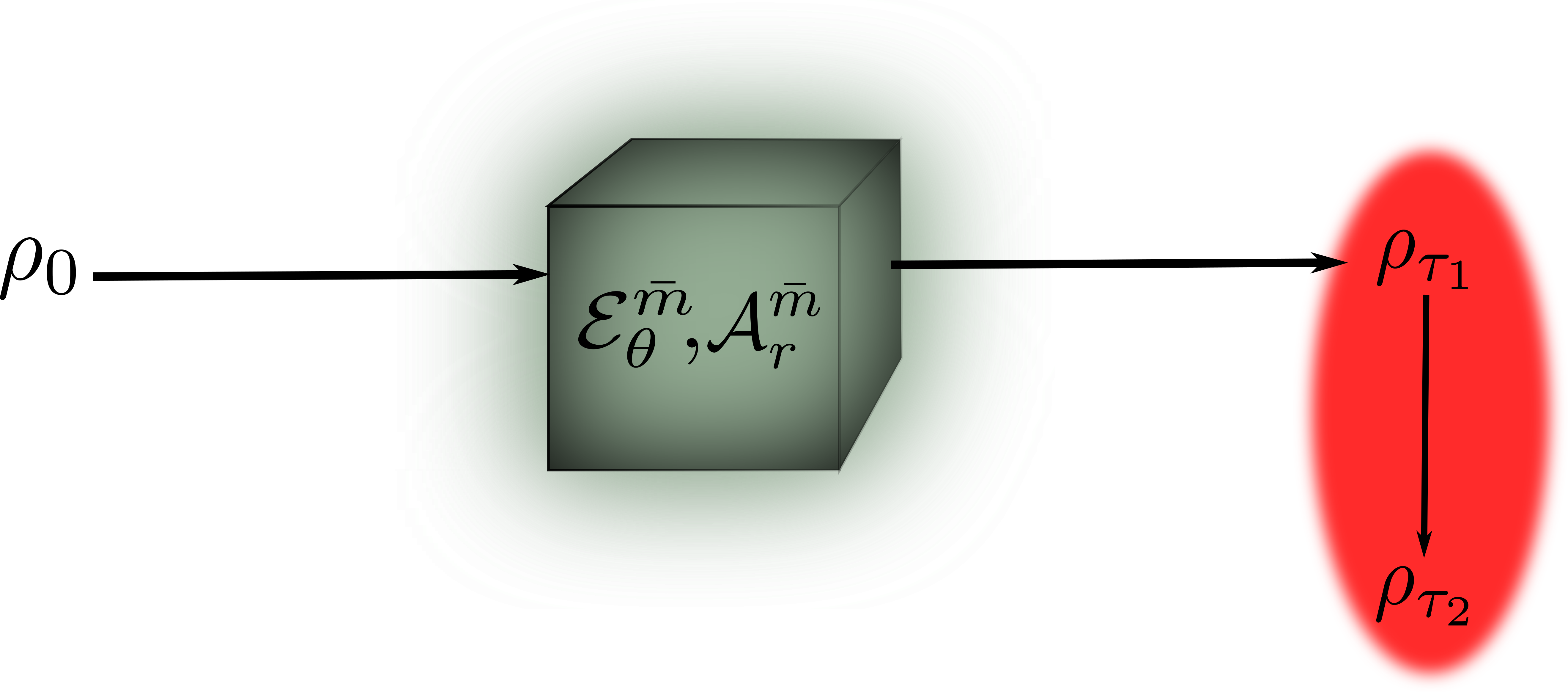}\caption{Entropy production. A single-mode input displaced thermal state $\rho_{0}$
passes through an attenuation (amplification) channel, which introduces
a coherence dynamics on the output state $\rho_{\tau_{1}}$. After
that, the system interacts with a Markovian heat reservoir and the
final state $\rho_{\tau_{2}}$ is obtained. Depending on the allocation
of the parameters $\theta$ ($r$), the coherence dynamics will result
in a larger or lower amount of entropy production.}

\label{ep_figure}
\end{figure}

Quantum thermodynamics is a modern topic that will impact many technological
areas such as quantum information science and quantum computation.
On one hand, coherence plays important roles in quantum thermodynamics
protocols. In particular, it is associated to the degradation or improvement
of the performance of quantum thermal machines \cite{Chen2017,Camati2019,Zagoskin2012},
as well as it is used as a resource to several types of dynamics \cite{Winter2016,Dana2017,Kamin2020}.
As we observed in Sec. \ref{sec:Coherence-in-attenuation}, depending
on the value of parameters $\theta$ or $r$, the output states can
have different amount of coherence for the attenuation or amplification
channels, respectively. On the other hand, the entropy production
is a relevant quantity in classical and quantum thermodynamics, mainly
because it is associated to the irreversibility of a given protocol
\cite{Jader2019,Gherardini2018,Deffner2011}. Besides, the increase
of the entropy production due to the presence of coherence has been
addressed in Ref. \cite{Jader2019,Francica2019}. Here we are interested
in relating the coherence dynamics dictated by an attenuation (amplification)
channel to the entropy production during a thermalization process.
With this purpose in mind the following protocol (see Fig. \ref{ep_figure})
is assumed. The system, prepared in a displaced thermal state with
average number of photons $\bar{n}$, interacts unitarily through
an attenuation (amplification) channel. This process is adiabatic
in the sense that there is no heat exchange. After that, the system
thermalizes with a Markovian thermal reservoir with average number
of photons $\bar{N}$ and inverse temperature $\beta=1/k_{B}T$, with
$k_{B}$ the Boltzmann constant. The entropy production in this case
can be written as (see Appendix \ref{sec:Appendix.-Entropy-production})

\begin{equation}
\langle\Sigma\rangle=-\beta\Delta\mathcal{U}_{\tau_{2},\tau_{1}}+\Delta S_{\tau_{2},\tau_{1}},\label{ep}
\end{equation}
where $\Delta\mathcal{U}_{\tau_{2},\tau_{1}}=\mathcal{U}_{\tau_{2}}-\mathcal{U}_{\tau_{1}}$
is the variation of internal energy associated to the thermalization
process, with $\mathcal{U}_{t}=\text{Tr\ensuremath{\left[\sigma_{t}\right]}}/4$,
and $\Delta S_{\tau_{2},\tau_{1}}=S(\rho_{\tau_{2}})-S(\rho_{\tau_{1}})$.
The relation between the inverse temperature and the average number
of photons of the reservoir is given by the Maxwell-Boltzmann distribution
$\bar{N}=\left(e^{\beta\hbar\omega}-1\right)^{-1}$.

The dynamics of the state during the thermalization is given by the
master equation \cite{PetruccioneBook,SerafiniBook}

\begin{equation}
\frac{d\rho}{dt}=\gamma\left(\bar{N}+1\right)\mathcal{D}\left[a\right]\rho+\gamma\bar{N}\mathcal{D}\left[a^{\dagger}\right]\rho,
\end{equation}
 with $\mathcal{D}[\mathcal{O}]=\mathcal{O}\rho\mathcal{O}^{\dagger}-(1/2)\left(\rho\mathcal{O}^{\dagger}\mathcal{O}+\mathcal{O}^{\dagger}\mathcal{O}\rho\right)$
the Lindblad operators and the decay rate $\gamma$. The same dynamics
can be recasted by acting on the first moments and covariance matrix
by employing the map $\mathcal{F}^{\text{therm}}=e^{-\gamma t/2}\mathbb{I}_{2}$
and $\mathcal{G}^{\text{therm}}=\left(2\bar{N}+1\right)\left(1-e^{-\gamma t}\right)\mathbb{I}_{2}$
in Eqs. (\ref{action1}) and (\ref{action2}) \cite{SerafiniBook}.

\begin{figure}
\includegraphics[scale=0.75]{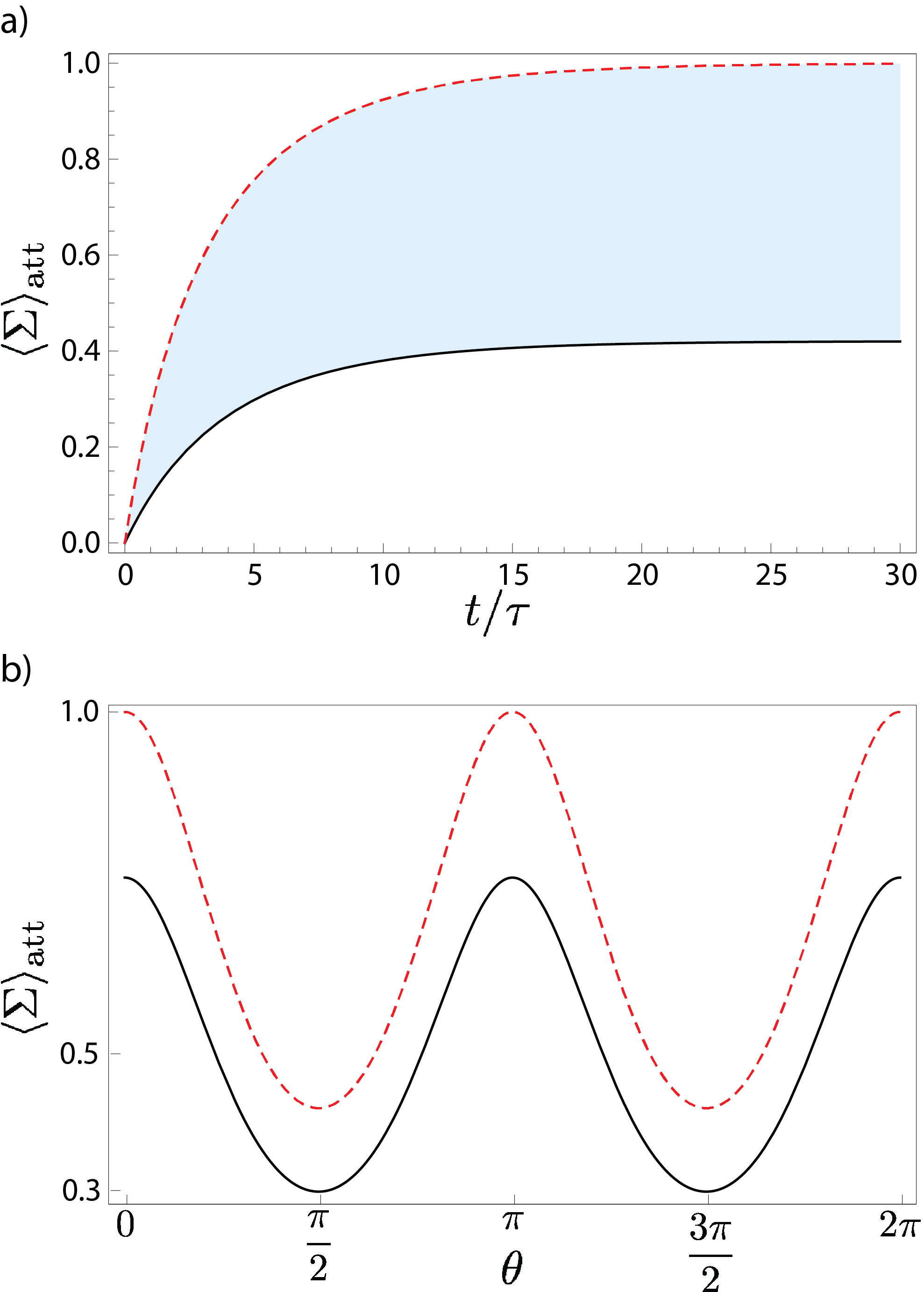}\caption{Entropy production for the attenuation channel. We assume the following
protocol: the input state $\rho_{\text{in}}$ with average number
of photons $\bar{n}$ passes through an attenuation channel characterized
by an angle $\theta$ and average number of photons $\bar{m}$; then
the output state $\rho_{\tau_{1}}$ thermalizes with a thermal reservoir
with average number of photons $\bar{N}$ such that $\bar{N}>\bar{n}$.
The final state of the thermalization is $\rho_{\tau_{2}}$. a) Entropy
production as a function of the time of interaction with the thermal
reservoir for $\theta=\pi/2$ (black solid line) and $\theta=0$ (red
dashed line). The parameter $\tau$ is introduced just to make the
time dimensionless. b) Entropy production as a function of the parameter
$\theta$ for $t/\tau=5$ (black solid line) and $t/\tau=T_{\text{th}}$
(red dashed line). We considered $\bar{n}=1$, $\bar{m}=2$, $\bar{N}=5$,
and $\gamma=0.1\tau$.}

\label{EPatt}
\end{figure}

\begin{figure}
\includegraphics[scale=0.75]{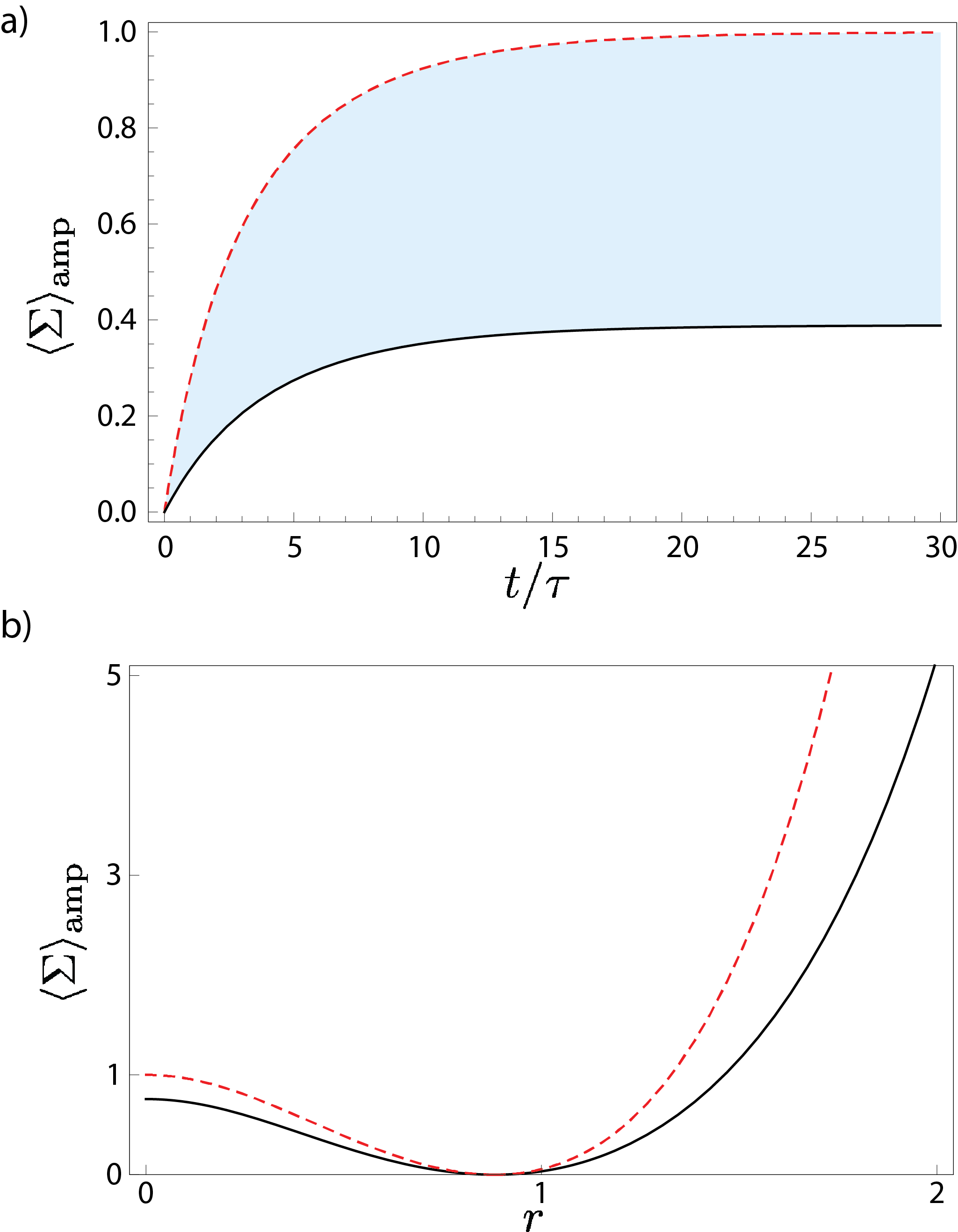}\caption{Entropy production for the amplification channel. We assume the following
protocol: the input state $\rho_{\text{in}}$ with average number
of photons $\bar{n}$ passes through an amplification channel characterized
by an squeezing parameter $r$ and average number of photons $\bar{m}$;
then the output state $\rho_{\tau_{1}}$ thermalizes with a thermal
reservoir with average number of photons $\bar{N}$ such that $\bar{N}>\bar{n}$.
The final state of the thermalization is $\rho_{\tau_{2}}$. a) Entropy
production as a function of the time of interaction with the thermal
reservoir for $r=0$ (red dashed line) and $r=0.5$ (black solid line).
The parameter $\tau$ is introduced just to become the time dimensionless.
b) Entropy production as a function of the parameter $r$ for $t=5/\tau$
(black solid line) and $t=T_{\text{th}}/\tau$ (red dashed line).
We consider $\bar{n}=1$, $\bar{m}=2$, $\bar{N}=5$, and $\gamma=0.1\tau$.}

\label{EPamp}
\end{figure}

To show how the coherence dynamics, introduced by the attenuation
(amplification) channel, affects the thermalization process, in Fig.
\ref{EPatt} (\ref{EPamp}) we present the entropy production as a
function of the parameter $\theta$ ($r$) and of the thermalization
time $t$ for the attenuation (amplification) channel, respectively.
For both situations we consider $\bar{n}=1$, $\bar{m}=2,$ $\bar{N}=5$,
and $\gamma=0.1\tau$. The parameter $\tau$ is introduced just to
make the time dimensionless. Figure \ref{EPatt}-a) shows the entropy
production for the attenuation channel as a function of the thermalization
time for $\theta=\pi/2$ (black solid lines) and $\theta=0$ (red
dashed lines). From our results for the coherence in this class of
channel, we note that the presence of coherence (dashed red line)
has an additional entropy production cost. On the other hand, Fig.
\ref{EPatt}-b) depicts the entropy production as a function of the
parameter $\theta$ fixing the thermalization time in $t/\tau=5$
(black solid line) (partial thermalization regime) and $t/\tau\approx T_{\text{th}}$(red
dashed line) (complete thermalization regime), with $T_{\text{th}}$
being the time need to have complete thermalization. As expected,
the values of $\theta$ for which there is no coherence ($\theta=\pi/2$
and $3\pi/2$) are those where the entropy production for complete
or partial thermalization reaches a minimum.

Figure \ref{EPamp}-a) illustrates the entropy production for the
amplification channel as a function of the thermalization time for
$r=0$ (dashed red line) and $r=0.5$ (solid black line). We observe
that for $r=0.5$ the entropy production is smaller than for $r=0$
for all time interval. This happens because the coherence, as we see
in Fig. \ref{cohereces1mode}-b), starts to decrease as $r$ becomes
larger up to reach an asymptotic value. In Fig.\ref{EPamp}-b) we
present the entropy production for the amplification channel as a
function of the parameter $r$ for $t/\tau=5$ (black solid line),
partial thermalization regime, and $t/\tau=T_{\text{th}}$ (red dashed
line), complete thermalization regime. It can be noted that, the entropy
production in both cases decreases up to a fixed value of $r$, when
then it starts to increase, and it goes to infinity as $r\rightarrow\infty$.
This change in the behavior of the entropy production is directly
associated to the coherence dynamics in \ref{cohereces1mode}-b).
We could argue that, for a given protocol where is expected to have
a minimum entropy production, the simulation through an amplification
channel could be useful, since the high control of the parameter $r$
is achievable.

From Figs. \ref{EPatt}-a) and .\ref{EPamp}-a) it is also possible
to conclude that the difference between the dashed red line and the
solid black line represents exactly the entropy production contribution
due to the coherence of the state $\rho_{\tau_{1}}$. In general,
we can express this mathematically as

\begin{equation}
\langle\Sigma\rangle_{\text{coherence}}^{\alpha}=\langle\Sigma\rangle(\ell_{1})-\langle\Sigma\rangle(\ell_{0}),
\end{equation}
where $\alpha$ stands for the attenuation (amplification) channel
and $\ell$ for the parameter $\theta$ ($r$), such that $\ell_{0}=\pi/2$
($r\rightarrow\infty)$ and $\ell_{1}=\theta$ ($r$), with $0\leq\theta\leq\pi/2$
and $0\leq r\leq\infty$. The hatched area in Figs. \ref{EPatt}-a)
and .\ref{EPamp}-a) means the quantity $\langle\Sigma\rangle_{\text{coherence}}^{\alpha}$
for different values of $\ell_{1}$. It must to be stressed that,
for the amplification channel, $r\rightarrow\infty$ represents the
minimum value of coherence that the output state $\rho_{\tau_{1}}$
can have, differently of the attenuation channel, in which the minimum
value of coherence for the output state is zero.

\section{Conclusion\label{sec:Conclusion}}

Gaussian channels play important role in continuous variables quantum
information and thermodynamics protocols. In particular, attenuation
and amplification channels are useful to describe the action of an
environment on a input state, as noise and decoherence. As Gaussian
operations, these channels can induce a coherence dynamics on a input
Gaussian state depending on the choice of the channels parameters.
In this work, we studied the how the attenuation and amplification
channels affect the initial coherence of a input Gaussian state. We
present analytical expressions for the coherence, showing that they
can vary considerably as a function of the channel structure.

The entropy production of an attenuation (amplification) operation
followed by an interaction with a Markovian thermal reservoir also
was considered. We have shown that, depending on the precise allocation
of the parameter in each type of channel, the entropy production can
be reduced, such that the implementation of some protocol in quantum
thermodynamics by means of the attenuation or amplification channels
may be useful in experimental realizations. In particular, for the
attenuation channel, it can be employed to simulate the transition
between quasi-static and finite-time regimes in driven systems. Finally,
for this class of channels we also write an expression relating the
entropy production cost due to the coherence dynamics of a output
state which thermalizes with a Markovian thermal reservoir. We hope
that this work can contribute with current and future experimental
propose to simulate thermodynamic process, as work extraction and
quantum thermal cycles employing continuous variable systems.

\section*{Acknowledgments}

Jonas F. G. Santos acknowledges São Paulo Research Grant No. 2019/04184-5,
for support. Carlos H. S. Vieira acknowledges CAPES (Brazil) for support.
The authors acknowledge Federal University of ABC for support.

\section*{Appendix. Entropy production\label{sec:Appendix.-Entropy-production}}

Although simple, here we derive the expression in Eq.(\ref{ep}).
As mentioned in main text, the protocol is composed of two parts.
The Gaussian input state passes through an attenuation (amplification)
channel and then thermalizes with a Markovian thermal reservoir with
inverse temperature $\beta$ and average number of photons $\bar{N}$.
The entropy production of a unitary process followed by a thermalization
can be written as \cite{Camati2019}

\begin{equation}
\langle\Sigma\rangle=S(\rho_{\tau_{1}}||\rho_{\tau_{1}}^{\text{eq,h}})-S(\rho_{\tau_{2}}||\rho_{\tau_{2}}^{\text{eq,h}}),
\end{equation}
 where $\rho_{\tau_{1}}$is the state after the unitary transformation,
$\rho_{\tau_{2}}$ is the state during the thermalization process,
and $\rho_{\tau_{1}}^{\text{eq,h}}$ is the reference thermal state
associated to the thermal reservoir. As the reference state is thermal
the relative entropy can be rewritten as $S(\rho_{t}||\rho_{t}^{\text{eq,h}})=\beta\left[\mathcal{U}(\rho_{t})-F_{t}^{\text{eq}}\right]-S(\rho_{t})$,
with $\mathcal{U}(\rho_{t})$ and $F_{t}^{\text{eq}}$ the internal
energy of the system and the free energy, respectively. By manipulating
we obtain

\begin{align*}
\langle\Sigma\rangle & =\beta\left[\mathcal{U}(\rho_{\tau_{1}})-F_{\tau_{1}}^{\text{eq}}\right]-S(\rho_{\tau_{1}})\\
 & -\left\{ \beta\left[\mathcal{U}(\rho_{\tau_{2}})-F_{\tau_{2}}^{\text{eq}}\right]-S(\rho_{\tau_{2}})\right\} \\
 & =\beta\left[\mathcal{U}(\rho_{\tau_{1}})-\mathcal{U}(\rho_{\tau_{2}})\right]-S(\rho_{\tau_{1}})+S(\rho_{\tau_{2}})\\
 & =-\beta\Delta\mathcal{U}_{\tau_{2},\tau_{1}}+\Delta S_{\tau_{2},\tau_{1}}.
\end{align*}

To complete, for Gaussian states the internal energy can be written
in terms of the covariance matrix, i.e.

\begin{equation}
\mathcal{U}(\rho_{t})=\frac{\hbar\omega}{4}\text{Tr}\left[\sigma_{t}\right].
\end{equation}
 With this, for Gaussian states the entropy production can be completely
characterized by the covariance matrix of the state.

\end{document}